\documentclass[journal,transmag]{IEEEtran}
\ifCLASSINFOpdf
  \usepackage[pdftex]{graphicx}
\else
\fi
\usepackage{amsmath}
\usepackage{siunitx}
\usepackage[caption=false]{subfig}
\usepackage{color}
\usepackage{amsfonts}
\usepackage{cancel}
\usepackage{booktabs}
\usepackage{multirow}
\usepackage{amssymb}
\usepackage{mathtools}
\newcommand{\adjustedaccent}[1]{%
\mathchoice{}{}
{\mbox{\raisebox{-.5ex}[0pt][0pt]{$\scriptstyle#1$}}}
{\mbox{\raisebox{-.35ex}[0pt][0pt]{$\scriptscriptstyle#1$}}}
}

\newcommand\bow[1]{\overset{\adjustedaccent{\smallfrown}}{#1}}

\begin{document}
%
\title{A Dynamic Energy-Based Hysteresis Model\\
for Pulsed-Operated Fast-Ramping Magnets}


\author{\IEEEauthorblockN{Dominik Moll\IEEEauthorrefmark{1},
Laura A. M. D'Angelo\IEEEauthorrefmark{1},
Herbert De Gersem\IEEEauthorrefmark{1},
Fulvio Boattini\IEEEauthorrefmark{2}, Luca Bottura\IEEEauthorrefmark{2}, 
Marco Gast\IEEEauthorrefmark{3}}
\IEEEauthorblockA{\IEEEauthorrefmark{1}Institute for Accelerator Science and Electromagnetic Fields,
TU Darmstadt, Germany}
\IEEEauthorblockA{\IEEEauthorrefmark{2}CERN, Geneva, Switzerland}
\IEEEauthorblockA{\IEEEauthorrefmark{3}Karlsruhe Institute for Technology, Germany}%
\thanks{Manuscript received July, 2025; 
Corresponding author: D. Moll (email: dominik.moll1@tu-darmstadt.de).}}

\markboth{Journal of \LaTeX\ Class Files,~Vol.~14, No.~8, July~2025}%
{Shell \MakeLowercase{\textit{et al.}}: Bare Demo of IEEEtran.cls for IEEE Transactions on Magnetics Journals}
%



\IEEEtitleabstractindextext{%
\begin{abstract}
Due to the strongly nonlinear
behavior of ferromagnetic yokes, the numerical analysis
of fast-ramping magnets is highly cumbersome
and, therefore, in practice overly simplified by means of anhysteretic material
descriptions and a posteriori loss formulae.
This paper establishes the use of a dynamic ferromagnetic model
combining a preconditioned energy-based hysteresis description
and a thin-sheet eddy-current model in time-domain.
The model was successfully employed in the analysis of a normal-conducting bending magnet
in order to
precisely calculate losses
and fields.
\end{abstract}

}

\maketitle

\IEEEdisplaynontitleabstractindextext

%
\IEEEpeerreviewmaketitle

\section{Introduction}
%
%
%
%
%
%

\IEEEPARstart{F}{ast}-ramping normal-conducting magnets delivering rise rates of several $\si{kT/s}$
receive increasing attention as 
they fill various key roles in
particle accelerator lattices.
A synchrotron-based muon collider will make extensive use
of such magnets in the intermediate acceleration stages~\cite{Bottura, Berg}.
The magnets will be excited by periodically pulsed currents
with very short duty cycles~\cite{Fulvio},
in order to minimize resistive loss
while reliably enforcing the requested ramp-up of the magnetic field.

The ramped excitation and the locally strong saturation of the ferromagnetic yoke
prohibit the application of frequency-domain approaches.
Moreover, calculating the hysteresis loss by post-processing an anhysteretic solution,
e.g. by the Steinmetz-Bertotti formulae~\cite{Bertotti}, yields inaccurate results
as overall energy conservation and the time-lag between exciting current
and generated magnetic field are neglected.
In order to precisely resolve the ferromagnetic behavior
of hysteresis and eddy-current phenomena,
an adequate dynamic material model as in~\cite{bib:gyselink1999, bib:zirka2004} will be formulated and applied.

In the following sections, the numerical representations of Maxwell's equation
and the applied dynamic energy-based hysteresis model are introduced.
The proposed method is then successfully applied on one of the current designs of a fast-ramping muon-collider magnet.
Lastly, the computational costs are compared and discussed.

\section{Formulation}
The magnetic vector potential approach resolves the magnetic law of Gauss and the law of Faraday-Lenz as
\begin{gather}
	\mathbf{B}=\nabla\times\mathbf{A},\\
	\mathbf{E}=-\dot{\mathbf{A}}-\nabla\varphi,
\end{gather}
with magnetic flux density $\mathbf{B}(\mathbf{r}, t)$, electric field strength $\mathbf{E}(\mathbf{r}, t)$,
magnetic vector potential $\mathbf{A}(\mathbf{r}, t)$
and electric scalar potential $\mathbf{\varphi}(\mathbf{r}, t)$.
With current density $\mathbf{J}=\sigma\mathbf{E}$ we formulate the law of Amp${\grave{\text{e}}}$re as
\begin{gather}
	\nabla\times\mathbf{H}+\sigma\dot{\mathbf{A}}=-\sigma\nabla\varphi,\label{ampere}
\end{gather}
with
$\mathbf{H}(\mathbf{r}, t)$ the magnetic field strength
and $\sigma(\mathbf{r})$ the conductivity.
A so-called field-circuit coupling is needed to account
for the external circuitry creating the pulse.
Every conductor is given a distribution function $\mathbf{x}_m(\mathbf{r})$~\cite{Schops} such that
\begin{gather}
-\sigma\nabla\varphi=\sum_{m}u_m\sigma\mathbf{x}_m,
\end{gather}
where $u_m(t)$ are the voltages along the conductors.
The currents through the solid conductors are given by
\begin{gather}
  	i_m=G_m u_m - (\mathbf{x}_m, \sigma\dot{\mathbf{A}})_{\Omega},\label{solid}
\end{gather}
where $G_m=(\mathbf{x}_m, \sigma \mathbf{x}_m)_\Omega$ is
the conductance of conductor $m$
and $(\cdot, \cdot)_\Omega$ denotes the $L^2$ scalar product over the computational domain $\Omega$.
The variational form of ($\ref{ampere}$) is
\begin{gather}
	(\mathbf{v},\mathbf{H})_\Omega+\cancelto{0}{(\mathbf{H}\times\mathbf{v},\mathbf{n})_{\partial\Omega}}+(\mathbf{w},\sigma\dot{\mathbf{A}})_\Omega
	=\sum_m u_m (\mathbf{w},\sigma\mathbf{x}_m)_{\Omega},
\end{gather}
where the test function $\mathbf{w}$ belongs to the space $H(\text{curl}, \Omega)$
of tangentially continuous vector fields with vanishing tangential components at the Dirichlet boundaries, and
$\mathbf{v}=\nabla\times\mathbf{w}$.
\subsection{Discretization}
Adopting the Ritz-Galerkin approach, i.e., using the same functions $\mathbf{w}_n$ for discretizing
$\mathbf{A}\approx\sum_n a_n(t)\mathbf{w}_n$
and testing one obtains
\begin{gather}
	\bow{\mathbf{h}}(\mathbf{a}, \dot{\mathbf{a}})+\mathbf{M}_{\sigma}\dot{\mathbf{a}}
	=\mathbf{X}\mathbf{u},\label{ampere_disc_voltage}
\end{gather}
with $\bow{h}_j=(\mathbf{v}_j, \mathbf{H})_\Omega$, $\mathbf{M}_{\sigma}|_{j,k}=(\mathbf{w}_j, \sigma\mathbf{w}_k)_\Omega$,
$\mathbf{X}|_{j,m}=(\mathbf{w}_j, \sigma\mathbf{x}_m)_\Omega$.
In case of known current excitations the system can be reformulated by incorporating~(\ref{solid})
\begin{gather}
	\bow{\mathbf{h}}(\mathbf{a}, \dot{\mathbf{a}})+(\mathbf{M}_{\sigma}-\mathbf{X}\mathbf{R}\mathbf{X}^\top)\dot{\mathbf{a}}
	=\mathbf{X}\mathbf{R}\mathbf{i},\label{ampere_disc_current}
\end{gather}
with $\mathbf{R}|_{j,m}=R_m\delta_{jm}$ as diagonal matrix of all resistances.

The right hand side of (\ref{ampere_disc_voltage}) (or selectively (\ref{ampere_disc_current})) is known for all times.
A Newton based solving scheme expands the left hand side locally around a guess $\mathbf{a}'$, $\dot{\mathbf{a}}'$
   \begin{gather}
	\bow{\mathbf{h}}(\mathbf{a}, \dot{\mathbf{a}})+\mathbf{M}_{\sigma}\dot{\mathbf{a}}
	\approx \nonumber\\
    \bow{\mathbf{h}}(\mathbf{a}', \dot{\mathbf{a}}')+\mathbf{M}_{\sigma}\dot{\mathbf{a}'}
       +\frac{\partial\bow{\mathbf{h}}'}{\partial\mathbf{a}}\delta\mathbf{a}
       +\left(\frac{\partial\bow{\mathbf{h}}'}{\partial\dot{\mathbf{a}}}+\mathbf{M}_\sigma\right)\delta\dot{\mathbf{a}},
\end{gather}
with $\delta\dot{\mathbf{a}}=\dot{\mathbf{a}}-\dot{\mathbf{a}}'$, $\delta{\mathbf{a}}={\mathbf{a}}-{\mathbf{a}}'$ and
\begin{gather}
	\frac{\partial\bow{h}_j}{\partial a_k}
	=\left(\mathbf{v}_j,\frac{\partial\mathbf{H} }{\partial a_k}\right)_\Omega
	=\left(\mathbf{v}_j,\frac{\partial\mathbf{H} }{\partial \mathbf{B}}\mathbf{v}_k\right)_\Omega,
    \nonumber
	\\
	\frac{\partial\bow{h}_j}{\partial \dot{a}_k}
	=\left(\mathbf{v}_j,\frac{\partial\mathbf{H} }{\partial \dot{a}_k}\right)_\Omega
	=\left(\mathbf{v}_j,\frac{\partial\mathbf{H} }{\partial \dot{\mathbf{B}}}\mathbf{v}_k\right)_\Omega.
    \nonumber
\end{gather}
Within the scope of this paper we will expect $\frac{\partial\mathbf{H} }{\partial \dot{\mathbf{B}}}$
to be constant, such that the system of differential equations is given in explicit form $\mathbf{M}\dot{\mathbf{a}}=\mathbf{f}(t,\mathbf{a})$
and standard time-integration routines can be applied.
We use the Python package \texttt{Pyrit}~\cite{Pyrit}
to assemble all finite element systems.

\section{Magnetic Constitutive Laws}
On the macroscopic scale the magnetic flux density and the magnetic field strength
are related via the magnetization $\mathcal{M}:\mathbf{H}\mapsto\mathbf{M}$
\begin{gather}
    \mathbf{B}=\mu_0(\mathbf{H}+\mathbf{M}(\mathbf{H}))=(\mu_0+\mu_0\mathcal{M})\mathbf{H}.
\end{gather}
In this section we characterize the ferromagnetic response of the
M235-35A soft-magnetic alloy, which will serve as
iron yoke material in the accelerator magnet design.
Extensive measurement data on this material can be found
and was used in the frame of this work~\cite{bib:Steentjes, bib:M235Estimate, bib:henrotte2013}.

\subsection{Anhysteretic Material Response}
\label{sec3a}
The \underline{an}hysteretic magnetization shall be given by an isotropic relationship as in~\cite{bib:Steentjes}
\begin{gather}
    |\mu_0\mathbf{M}_\text{an}|=\mu_{0}M_a L(\frac{|\mathbf{H}|}{h_a})+\mu_{0}M_b L(\frac{|\mathbf{H}|}{h_b}),
\end{gather}
with Langevin function $L(x)=\coth(x)-x^{-1}$, $\mu_{0}M_a=\SI{1.39}{T}$, $h_a=\SI{18.18}{A/m}$, $\mu_{0}M_b=\SI{0.56}{T}$, $h_b=\SI{3.91}{kA/m}$.
The maximum susceptibility of $\chi_{\max} ={20.32}\,\,10^{3}$
is found for very small excitations $|\mu_0\mathbf{M}_\text{an}|=\mu_{0}\chi_{\max} |\mathbf{H}|+\mathcal{O}(|\mathbf{H}|^3)$.

Anhysteretic models are commonly applied as they offer a rapid evaluation of $\mathbf{H}\mapsto\mathbf{B}$
and also $\mathbf{B}\mapsto\mathbf{H}$, whereby the latter may require the use of look-up-tables
as analytical inverses often can not be provided.
Similarly straight forward is the construction of the differential tensors
\begin{gather}
    \frac{\partial\mathbf{B}}{\partial\mathbf{H}}=
    \frac{|\mu_0\mathbf{M}_\text{an}|}{|\mathbf{H}|}[\mathbf{I}-\mathbf{e}_\mathbf{H}\mathbf{e}^\top_\mathbf{H}]
    +\frac{\partial |\mu_0\mathbf{M}_\text{an}|}{\partial |\mathbf{H}|}\mathbf{e}_\mathbf{H}\mathbf{e}^\top_\mathbf{H}.
\end{gather}
By construction, this type of material model is incapable of
reproducing the hysteretic behavior of ferromagnetic alloys.
Accordingly, the hysteresis loss can only be approximated using a posteriori formulae
for which we here assume the principle of loss separation~\cite{Bertotti}
combined with signal-independent generalized loss expressions~\cite{bib:venkata}.
The time-averaged hysteresis and eddy current loss densities are
\begin{gather}
	\bar{p}_\text{hyst}=\frac{\int_0^T \gamma k_{\text{hyst}}|\dot{\mathbf{B}}| \hat{B}\text{d} t/T}{\int_0^{2\pi} |2\pi\cos(\theta)|\text{d}\theta/2\pi},\\
	\bar{p}_\text{eddy}=\frac{\int_0^T \gamma k_{\text{eddy}}|\dot{\mathbf{B}}|^2 \text{d} t/T}{\int_0^{2\pi} |2\pi\cos(\theta)|^2\text{d}\theta/2\pi},
\end{gather}
with mass-density $\gamma$, empirically obtained parameters $k_{\text{hyst}}$, $k_{\text{eddy}}$
and $\hat{B}$ as one half of the peak to peak value of the local magnetic flux density.
The values $k_\text{hyst}=\SI{13.88}{mW/(kg Hz T^2)}$ and $k_\text{eddy}=\SI{44.77}{\micro W/(kg Hz^2 T^2)}$
have been reported in \cite{bib:M235Estimate} and are used here.
The so-called excess loss will not be discussed in this work.

\subsection{Hysteretic Material Response -- Forward}
\label{sec2c}
The energy-based hysteresis model of Bergqvist~\cite{bib:Bergqvist} and refined by Henrotte~\cite{bib:Henrotte2006}
expresses the magnetization as
\begin{gather}
\mathbf{M}(\mathbf{H})=\sum_k \mathbf{M}_k(\mathbf{H})=\sum_{k}{w_k}\mathcal{M}_{\text{an}}\mathcal{H}_{\text{r}}^k\mathbf{H},
\end{gather}
for which the weights $w_k$ fulfill $\sum_{k}{w_k}=1$.
The operator $\mathcal{M}_{\text{an}}$ realizes the anhysteretic mapping as defined in~\ref{sec3a}. 
A~history dependence is included in the operators $\mathcal{H}_{\text{r}}^k$
which can be approximated by a vector-play model with pinning forces $\kappa_k\geq 0$ (Tab.~\ref{tab:m235})
and Heaviside function $\Theta$ \cite{bib:Henrotte2006}
\begin{gather}
	\mathbf{H}_r^k=\mathcal{H}_r^k\mathbf{H}=\mathbf{H}_{r, \text{prev}}^k+f(|\delta\mathbf{H}_k|, \kappa_k)\mathbf{e}_{\delta\mathbf{H}_k},\\
	f(x, y)= (x-y)\,\Theta(x-y),\nonumber\\
	\delta\mathbf{H}_k=\mathbf{H}-\mathbf{H}_{r, \text{prev}}^k.\nonumber
\end{gather}
\begin{table}[b]
	\caption{Weights and pinning forces of M235-35A}
		\label{tab:m235}
	\centering
\begin{tabular}{rrr}
        \toprule
     $m$ (-) & $w_m$ (-) & $\kappa_m$ ($\si{A/m}$)\\
    \midrule
1 &0.07548	&0\\
2 &0.10322	&7.34865\\
3 &0.10637	&18.82524\\
4&	0.34187	&32.11778\\
5&	0.11947	&45.51681\\
6& 0.10531	&55.76191\\
7& 0.05298	&66.86223\\
8& 0.04347	&80.55601\\
9& 0.02820	&99.10729\\
10& 0.01931	&143.04169\\
11 &0.00551	&213.50904\\
    \bottomrule
\end{tabular}
	\end{table}
For unidirectional fields this approximation is fully consistent with the
energy conserving framework.
For arbitrary fields, errors are introduced by this model which can only
be diminished by solving a computationally more expensive optimization problem~\cite{bib:prigozhin2016}.

For both approaches the computation of the differential susceptibility tensor $\frac{\partial\mathbf{M}}{\partial\mathbf{H}}$
and differential permeability tensor $\frac{\partial\mathbf{B}}{\partial\mathbf{H}}=\mu_0+\mu_0\frac{\partial\mathbf{M}}{\partial\mathbf{H}}$
becomes a concatenation of analytical expressions~\cite{bib:jaquesdirectinverse}
\begin{gather}
\frac{\partial\mathbf{M}}{\partial\mathbf{H}}
=\sum_k w_k \frac{\partial\mathbf{M}_\text{an}}{\partial\mathbf{H}_\text{r}^k}\frac{\partial\mathbf{H}_\text{r}^k}{\partial\mathbf{H}}.
\end{gather}

The local power density $\mathbf{H}\cdot\dot{\mathbf{B}}$ translates into a change in stored magnetic energy $w_\text{mag}$
and a hysteretic loss
$p_{\text{hyst}}$
\begin{gather}
	\dot{w}_{\text{mag}}=\mathbf{H}\cdot\mu_0\dot{\mathbf{H}}+\sum_m\mathbf{H}_r^m\cdot\mu_0\dot{\mathbf{M}}_m,\\
	p_{\text{hyst}}=\sum_m\left(\mathbf{H}-\mathbf{H}_r^m\right)\cdot\mu_0\dot{\mathbf{M}}_m\geq 0.\label{EBmodelineq}
\end{gather}
Within the scope of this work the vector-play model was selected,
as the impact of the thereby introduced errors has been observed to be marginal
for our intended application of accelerator magnets.
\subsection{Hysteretic Material Response -- Inverse}
\label{sec3c}
The introduced vector-potential based solving algorithm~\eqref{ampere_disc_voltage}
requires the inverse mapping $(\mu_0+\mu_0\mathcal{M})^{-1}:\mathbf{B}\mapsto\mathbf{H}$.
A direct realization of this operation has been recently proposed~\cite{bib:egger2025}.
State-of-the-art implementations rely on fixed-point iteration
schemes based upon the forward model~\cite{bib:jaquesdirectinverse}.

The fixed-point iteration schemes for a given $\mathbf{B}^*$ are
characterized by $\mathcal{Q}:\mathbf{H},\mathbf{B}^*\mapsto \mathbf{Q}$ such that $\mathbf{H}_\text{n+1}=\mathbf{Q}(\mathbf{H}_n,\mathbf{B}^*)$
constructs a converging series with $\lim_{n\to\infty}\mathbf{H}_n=\mathbf{H}^*$ and $\mathbf{B}^*=\mathbf{B}(\mathbf{H}^*)$.
Root finding algorithms applied on the residual
\begin{gather}
\mathbf{g}(\mathbf{H}^*, \mathbf{B}^*)=\mathbf{B}(\mathbf{H}^*)-\mathbf{B}^*=\mathbf{0}
\end{gather}
lead to the following iterative schemes
\paragraph{Direct Iteration}
\begin{gather}
    \mathbf{Q}(\mathbf{H}, \mathbf{B}^*)=\mathbf{H}-(\mu_0\mu_{r,\max})^{-1}\mathbf{g}(\mathbf{H}, \mathbf{B}^*),
\end{gather}
\paragraph{Newton Iteration}
\begin{gather}
    \mathbf{Q}(\mathbf{H}, \mathbf{B}^*)
    =\mathbf{H}-\frac{\partial \mathbf{B}}{\partial \mathbf{H}}^{-1}\mathbf{g}(\mathbf{H},\mathbf{B}^*).
\end{gather}
The direct iteration has ensured convergence for all initial guesses $\mathbf{H}_0$ with $\mu_0\mu_{r, \max}$ being the maximum differential permeability
but converges slowly.
The Newton iteration is able to provide a much faster convergence if the initial guess is
well justified for example by using the previous result in combination with a very fine time stepping.
As the validity of the initial guess is not a priori known
the Newton method generally needs to be stabilized by additional success criteria and relaxation factors.
Furthermore, the differential tensors must be constructed for every iteration step.

To overcome the disadvantages of both of these well studied schemes
we introduce the preconditioned iteration scheme.
The preconditioned residual is defined as
\begin{gather}
\mathbf{g}_p(\mathbf{H}^*, \mathbf{B}^*)=\mathbf{B}_\text{an}^{-1}(\mathbf{B}(\mathbf{H}^*))-\mathbf{B}_\text{an}^{-1}(\mathbf{B}^*)=\mathbf{0}
\end{gather}
yielding the
\paragraph{Preconditioned Iteration}
\begin{gather}
    \mathbf{Q}(\mathbf{H}, \mathbf{B}^*)=\mathbf{H}-\mathbf{g}_p(\mathbf{H}, \mathbf{B}^*)
\end{gather}
The preconditioned iteration thus actively exploits information on the anhysteretic magnetization curve $\mathbf{B}_\text{an}$
for which the inverse operation can be cheaply performed using look-up-tables.
It yields ensured convergence for the ferromagnetic material
and drastically reduces the amount of iteration steps as opposed to the direct iteration.

\begin{figure}[!t]
   \centering
   \includegraphics[width=0.9\columnwidth]{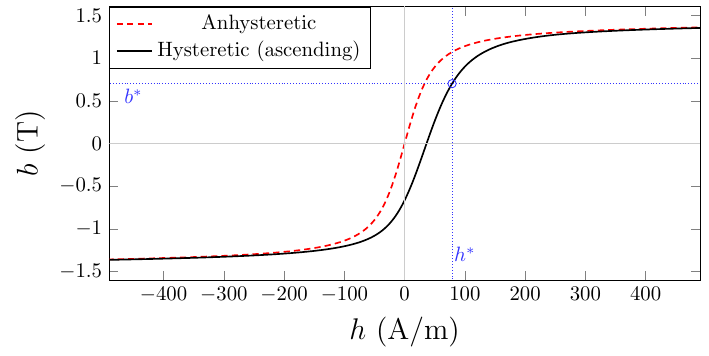}
   \caption{Anhysteretic and hysteretic material response (ascending branch, major cycle). Input $b^*$ and target $h^*$
       of the one-dimensional inverse hysteresis problem.}
   \label{fig:hyst1d}
\end{figure}

\begin{table}[t]
    \caption{Computation time in \si{\micro s} per problem and iteration count (bold)}
    \label{tab_iter}
  \begin{tabular}{crrrrrr}
    \toprule
    \multirow{2}{*}{relative error} &
      \multicolumn{3}{c}{$h_0=\SI{100}{A/m}$} &
      \multicolumn{2}{c}{$h_0=\SI{1}{kA/m}$} \\
      & {Direct} & {Newton} & {Precond.} & {Direct}  & {Precond.} \\
      \midrule
    \multirow{2}{*}{$10^{-3}$} & {13.2} &  10.2& 6.3 & 53.0  &7.2 \\
    &\textbf{10} & \textbf{3} & \textbf{4} & \textbf{48}  & \textbf{5} \\
    \multirow{2}{*}{$10^{-6}$} & 25.8 & 11.8 & 10.4 & 64.0  & 11.3 \\
    &\textbf{22} & \textbf{4} & \textbf{8} & \textbf{59}  & \textbf{9} \\
    \multirow{2}{*}{$10^{-9}$} & 38.0 & 11.8 & 15.2 & 77.1 & 16.4 \\
    &\textbf{33} & \textbf{4} & \textbf{13} & \textbf{70}  & \textbf{14} \\
    \bottomrule
  \end{tabular}
\end{table}
Identical one-dimensional inverse problems with $\mathbf{B}^*=\SI{0.7}{T}\,\mathbf{e}_\theta$
on the ascending branch of the major hysteresis cycle (Fig.~\ref{fig:hyst1d})
with angular direction ${\mathbf{e}_\theta=\cos(\theta)\mathbf{e}_x+~\sin(\theta)\mathbf{e}_y}$
have been solved for 36000 distinct angles using a serial implementation.
Table~\ref{tab_iter} displays the averaged runtime per problem and iteration count.
The solution is approximately given by $\mathbf{H}^*\approx\SI{78.68}{A/m}\,\mathbf{e}_\theta$.
For a comparably good initial guess of $\mathbf{H}_0=h_0\mathbf{e}_\theta=\SI{100}{A/m}\,\mathbf{e}_\theta$ the Newton iteration
performs best in terms of iterations needed and computation time for a strict tolerance of $10^{-9}$.
For lower tolerances the preconditioned iteration was found to be the fastest scheme.
Generally the preconditioned iteration has a time consumption
in the same order of magnitude as the Newton scheme.
Both methods clearly outperform the direct iteration.
For the significantly worse initial guess of $h_0=\SI{1}{kA/m}$
the iteration count of the preconditioned iteration only rises by one.
The Newton method is unable to converge and the time consumption of the direct iteration
scheme more than doubles.

The preconditioned iteration scheme describes a computationally inexpensive and stable
procedure to realize the inverse
hysteretic mapping $\mathbf{B}\mapsto\mathbf{H}$.
Unlike the Newton method, it does not necessitate relaxation criteria or sufficiently fine time stepping.
Based on these considerations we have concluded to use the
preconditioned iteration scheme of up to 20 iteration steps in this work.

\subsection{Dynamic Ferromagnetic Response}
\label{sec3d}
The magnetic response of soft-magnetic alloys
is further governed by
the eddy current problem
which in case of purely transversal fields with $\partial_x=\partial_y=0$ reduces to the so-called
one-dimensional lamination problem
\begin{gather}
-\partial_z^2\mathbf{H}=-\sigma_\text{Fe}\dot{\mathbf{B}}.
\end{gather}

It was shown that the dynamic model
\begin{gather}
    \mathbf{H}_{\text{surf}}=\mathbf{H}(\mathbf{B})+\frac{\sigma_\text{Fe} d^2}{12}\dot{\mathbf{B}}\label{dyna_BH}
\end{gather}
is able to approximate the magnetic field strength $\mathbf{H}_\text{surf}$ at the iron-insulation-interface
very well for low frequent periodic excitations~\cite{bib:gyselink1999, PDular}.
The eddy current loss is then given by
\begin{gather}
{p}_{\text{eddy}}=\frac{\sigma_\text{Fe} d^2}{12}|\dot{\mathbf{B}}|^2\geq 0.
\end{gather}
Further refinements of this dynamic model including excess terms have been suggested
but are not considered in this scope~\cite{bib:zirka2004}.

In the following section we will apply the anhysteretic and hysteretic material responses
with and without the eddy-current contribution of (\ref{dyna_BH}).

\section{Application}
\subsection{Magnet Layout}
\begin{figure}[h]
   \centering
   \includegraphics[width=0.75\columnwidth]{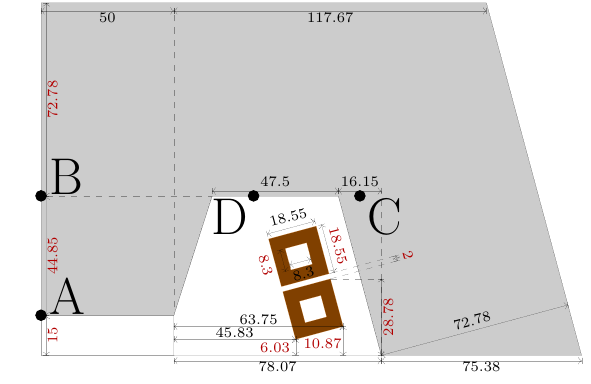}
   \caption{One quarter of the analyzed H-type dipole magnet. All lengths in millimeters.}
   \label{fig:magnet}
\end{figure}

A dipole magnet consisting of an angled H-type
ferromagnetic yoke and eight solid conductors (Fig.~\ref{fig:magnet}) will be analyzed.
This particular design is the result of a cost minimizer
for the second rapid cycling synchrotron (RCS2) of a potential muon collider~\cite{muoncollider_report2024}.
The provided dipole field must be
ramped from negative (\SI{-1.8}{T}) to positive peak (\SI{1.8}{T}) within $\SI{1}{ms}$~\cite{muoncollider_report2024, bib:breschi2024}.
The length measures several meters, such that end effects are of no interest in this initial study
and the assumption of purely transversal fields holds (\ref{sec3d}).

\subsection{Sinusoidal Excitation}
We firstly select a sinusoidal current excitation of $I=\SI{12.5}{kA}\sin(2\pi\cdot \SI{500}{Hz}\cdot t)$
per conductor
and compare the results of the discussed material models.
We simulate three periods with 100 equidistant
implicit Euler steps per $\SI{1}{ms}$ time intervall.

\begin{figure}[!t]
   \centering
   \includegraphics*[width=1\columnwidth]{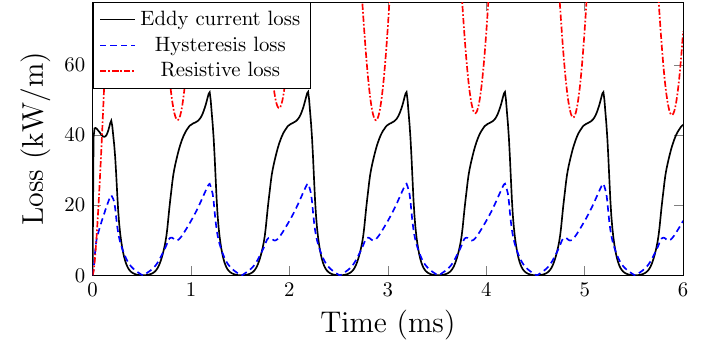}
   \caption{Loss as obtained by using the dynamic hysteresis model for a sinusoidal excitation.}
   \label{fig:loss_sin}
\end{figure}

The loss of the third cycle is given by $\SI{336.08}{J/m}$ and is heavily dominated by the
resistive loss occurring in the copper conductors (Tab.~\ref{tab_loss_sin} and Fig.~\ref{fig:loss_sin}).
The iron yoke loss contributes less than 20\% of the total loss
whereby the eddy current loss is more than twice as large as the hysteresis loss.
\begin{table}[t]
    \caption{Loss per cycle in $\si{J/m}$ for a sinusoidal excitation}
    \label{tab_loss_sin}
  \begin{tabular}{cccc}
    \toprule
       Material model  & {Eddy current} & {Hysteresis} & {Resistive} \\
      \midrule
    Anhysteretic (static) & 47.15 (a post.) &  23.54 (a post.) & 272.90 \\
    Anhysteretic (dynamic) & 41.22 &  0 & 272.88 \\
    Hysteretic (static) & 0 &  20.12 & 273.41 \\
    Hysteretic (dynamic) & 42.61 &  20.09 & 273.38 \\
    \bottomrule
  \end{tabular}
\end{table}

\begin{figure}[!t]
    \captionsetup[subfloat]{farskip=2pt,captionskip=1pt}
   \centering
   \subfloat[BH in point A.]{\includegraphics[width=0.49\columnwidth]{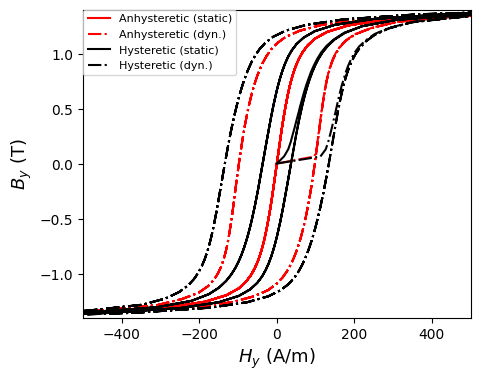}\label{fig:b_sin_1}}
   \hfill
   \subfloat[BH in point B.]{\includegraphics[width=0.49\columnwidth]{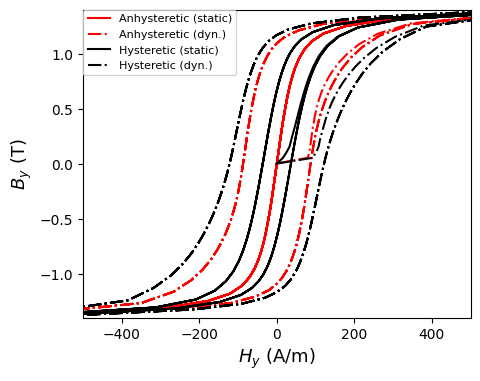}\label{fig:b_sin_2}}
   \\
   \subfloat[BH in point C.]{\includegraphics[width=0.49\columnwidth]{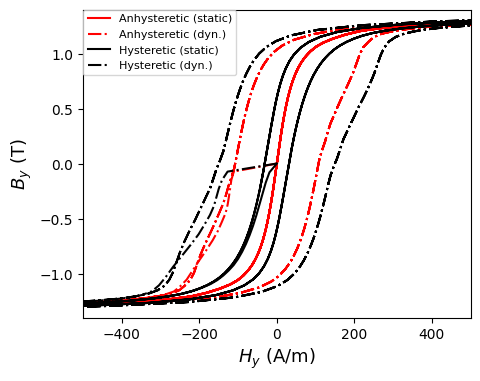}\label{fig:b_sin_3}}
   \hfill
   \subfloat[BH in point D.]{\includegraphics[width=0.49\columnwidth]{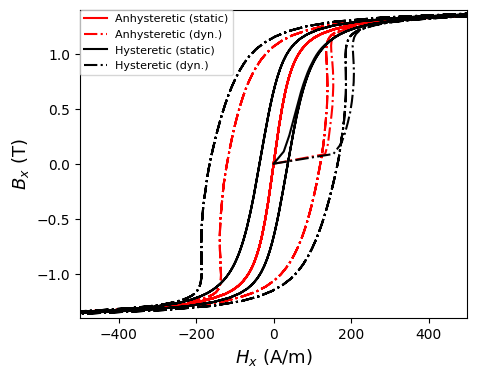}\label{fig:b_sin_4}}
	\caption{BH loci in the ferromagnetic yoke. Static model: solid lines. Dynamic model: dashed lines.}
          \label{fig:b_sin_0}
\end{figure}
\begin{figure}[!t]
   \centering
  \subfloat[Dipole component.]{\includegraphics[width=0.49\columnwidth]{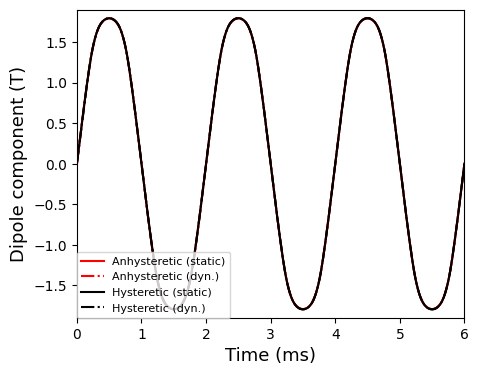}\label{fig:b_sin_5}}
   \hfill
     \subfloat[Relative differences of dipole component compared to the dynamic hysteresis model.]
     {\includegraphics[width=0.49\columnwidth]{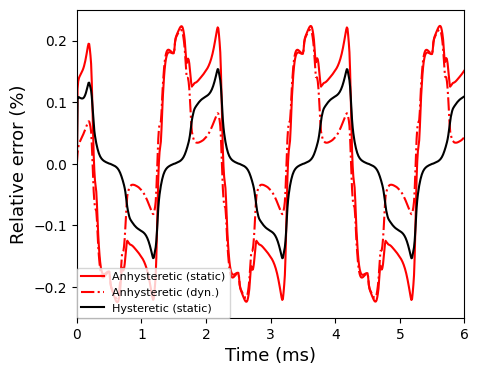}\label{fig:b_sin_6}}
	\caption{Dipole component in the air gap center.}
          \label{fig:b_sin_dip}
\end{figure}
All used ferromagnetic models agree in terms of resistive loss.
The eddy current and hysteresis loss obtained via the a~posteriori formulae is
11\% and 17\% larger than with the dynamic hysteresis model.
The discrepancy is caused by the a~posteriori loss factors $k_\text{eddy}$ and $k_\text{hyst}$
which are not tuned for the comparably high frequency of $\SI{500}{Hz}$~\cite{bib:M235Estimate}.
Our results underline the weaknesses of a posteriori loss estimates
and would allow to readjust $k_\text{eddy}$ and $k_\text{hyst}$ for this particular frequency.

The BH loci in four yoke points are shown in Fig.~\ref{fig:b_sin_0}.
The initial state is given by zero.
The static loci in all points are almost identical, whereas
the dynamic model causes a notably different broadening of the BH loop of each point.

The dipole component in the air gap center (Fig.~\ref{fig:b_sin_5}) is
not significantly influenced by the ferromagnetic model.
Relative differences in the order of $10^{-3}$ are observed (Fig.~\ref{fig:b_sin_6}).

\subsection{Triangular Pulsed Excitation}
\begin{figure}[h]
   \centering
   \includegraphics*[width=1\columnwidth]{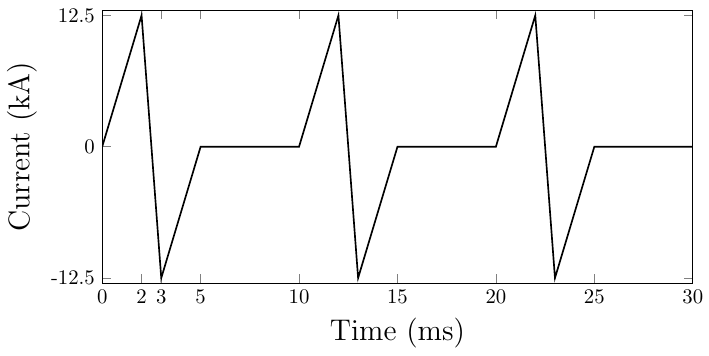}
   \caption{Triangular pulsed current. The fall rates are four times larger than the rise rates.}
   \label{fig:current_tri}
\end{figure}

We will now assume an excitation
of repeating triangular pulses (Fig.~\ref{fig:current_tri}) as idealized current supply of the magnet
obtained from a switched resonance circuit \cite{Fulvio}.
In the simulations the intended repetition rate of $\SI{5}{Hz}$ was increased to $\SI{100}{Hz}$
in order to shorten the constant zero-phases and thus reduce computational efforts.
Again 100 equidistant time steps per \SI{1}{ms} are applied.

\begin{figure}[!t]
   \centering
   \includegraphics*[width=1\columnwidth]{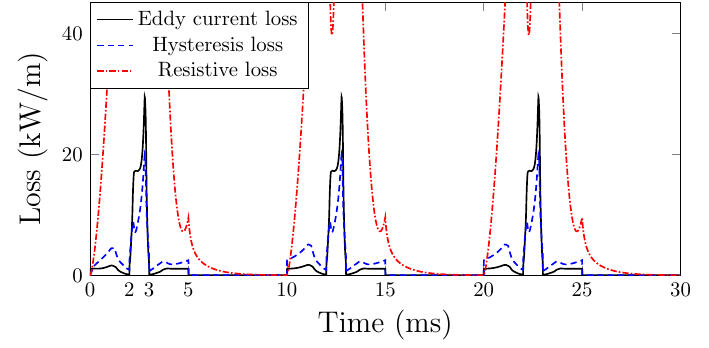}
   \caption{Loss as obtained by using the dynamic hysteresis
model for a pulsed excitation.}
   \label{fig:loss_tri}
\end{figure}

The hysteresis loss of the pulsed excitation is identical to the
result obtained for the sinusoidal excitation (Tab.~\ref{tab_loss_pulse}).
The eddy current loss is more than halved due to
the lower rise and fall rates.
The overall loss of one pulse with $\SI{358.70}{J/m}$ is however higher than for a single
sinusoidal period due to an increase of almost $\SI{46}{J/m}$
in resistive loss.
It is important to note that the resistive loss is not immediately
zero once a pulse terminates but instead decays exponentially (Fig.~\ref{fig:loss_tri}).

\begin{figure}[!t]
    \captionsetup[subfloat]{farskip=2pt,captionskip=1pt}
   \centering
      \subfloat[BH in point A.]{\includegraphics[width=0.49\columnwidth]{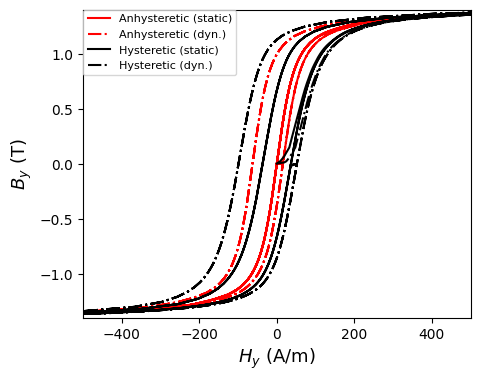}\label{fig:b_pulse1}}
   \hfill
  \subfloat[BH in point B.]{\includegraphics[width=0.49\columnwidth]{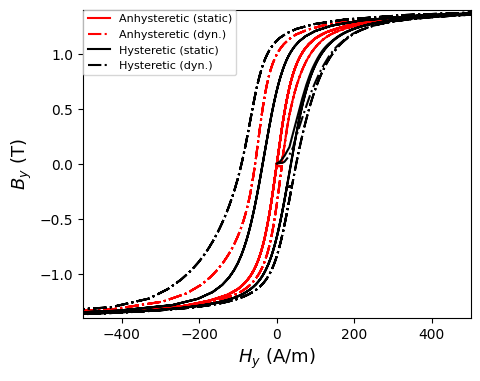}\label{fig:b_pulse2}}
\\
  \subfloat[BH in point C.]{\includegraphics[width=0.49\columnwidth]{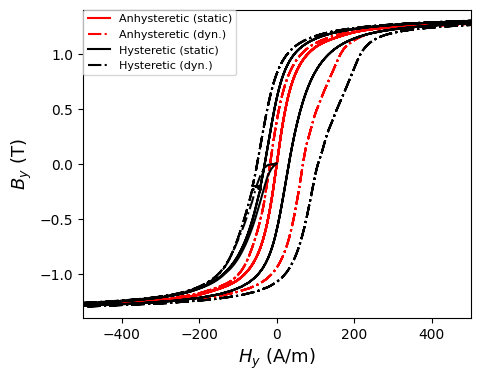}\label{fig:b_pulse3}}
   \hfill
  \subfloat[BH in point D.]{\includegraphics[width=0.49\columnwidth]{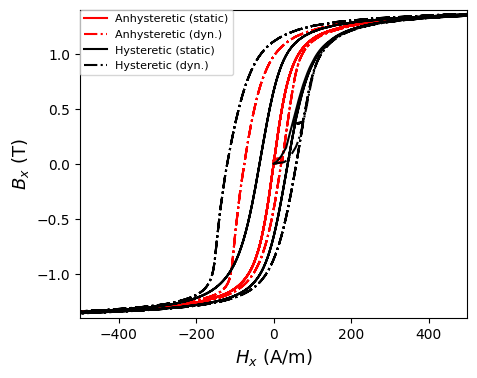}\label{fig:b_pulse4}}
	\caption{BH loci in the ferromagnetic yoke. Static model: solid lines. Dynamic model: dashed lines.}
  \label{fig:b_pulse0}
\end{figure}

The BH loci of the dynamic models (Fig.~\ref{fig:b_pulse0})
are narrower than for the sinusoidal excitation.
The ascending branch encloses a much smaller area
than the descending branch, due to significantly different rise and fall rates
of the current excitation.
We can furthermore identify plateaus at which
$\mathbf{B}$ and $\mathbf{H}$ transfer from dynamic to static behavior once a pulse terminates, such as
$B_x=\SI{0.4}{T}$ in point D (Fig.~\ref{fig:b_pulse4}).

\begin{figure}[!t]
   \centering
  \subfloat[Dipole component.]{\includegraphics[width=0.49\columnwidth]{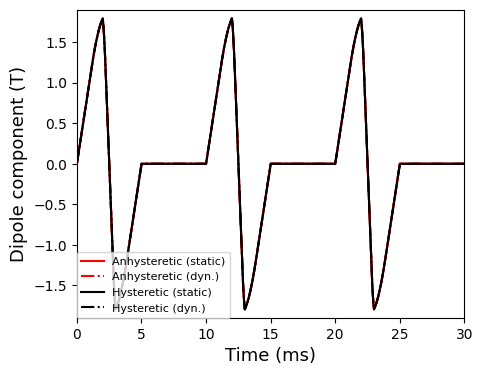}\label{fig:b_pulse5}}
   \hfill
     \subfloat[Relative differences of dipole component compared to the dynamic hysteresis model.]
     {\includegraphics[width=0.49\columnwidth]{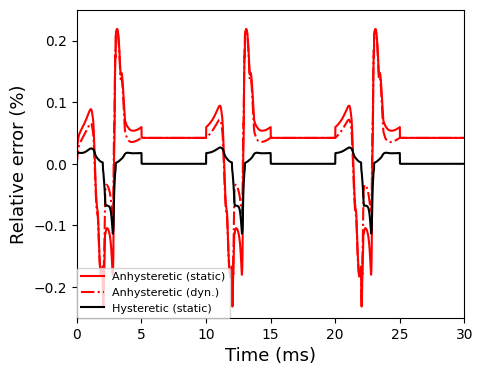}\label{fig:b_pulse6}}
	\caption{Dipole component in the air gap center.}
          \label{fig:b_pulse_dip}
\end{figure}

The resulting aperture field (Fig.~\ref{fig:b_pulse_dip})
again differs by only $0.2\%$.
The most noteworthy difference is the constant remanence field of $\SI{-0.75}{mT}$
after each pulse which can only be resolved by the hysteretic models.

\begin{table}[!t]
    \caption{Loss per cycle in $\si{J/m}$ for a pulsed excitation}
    \label{tab_loss_pulse}
  \begin{tabular}{cccc}
    \toprule
       Material model  & {Eddy current} & {Hysteresis} & {Resistive} \\
      \midrule
    Anhysteretic (static) & 21.63 (a post.) &  23.60 (a post.) & 318.79 \\
    Anhysteretic (dynamic) & 18.62 &  0 & 318.78 \\
    Hysteretic (static) & 0 &  20.14 & 319.23 \\
    Hysteretic (dynamic) & 19.38 &  20.10 & 319.22 \\
    \bottomrule
  \end{tabular}
\end{table}


\newpage
\section{Runtime Comparison}
The simulation runtime
increases linearly
with the timesteps~$N_t$,
nonlinear iteration steps $N_\text{Newton}$
and degrees of freedom~$N_\text{Nodes}$, i.e.,
$\mathcal{O}(N_tN_\text{Newton}N_\text{Nodes})$.
The calculations with anhysteretic and hysteretic model
only differ in the ferromagnetic material evaluation,
which is required in all $N_\text{int}^\text{Fe}$ integration points of the
iron yoke.
The additional time consumption related to
the serialized material
evaluation scales as
$\mathcal{O}(N_tN_\text{Newton}N_\text{int}^\text{Fe})$.

\begin{figure}[!h]
   \centering
   \includegraphics*[width=0.95\columnwidth]{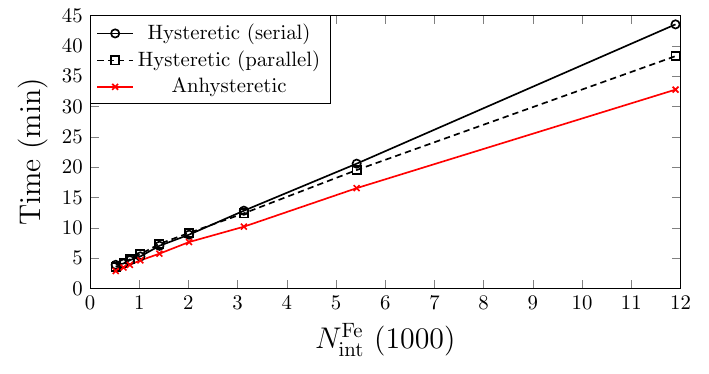}
   \caption{Runtime for solving the transient FE problem with sinusoidal excitation on uniformly resolved meshes.
   The parallel material evaluation used two threads.}
   \label{fig:runtime}
\end{figure}

The calculation with sinusoidal excitation
was repeated for various uniformly resolved meshes using a standard desktop computer with
{10}\,{cores} and {32}\,{GB} of RAM.
The amount of time steps and Newton iterations was
fixed as $N_t=600$ and $N_\text{Newton}=5$.
The simulation with a mesh of $N_\text{int}^\text{Fe}=11898$ integration points
required \SI{32.7}{min} with anhysteretic materials
and \SI{43.5}{min} with hysteretic materials (Fig.~\ref{fig:runtime}).
The difference of $\SI{10.8}{min}$ translates into an additional
individual
evaluation time of \SI{18.2}{\micro s}
per integration point which is in a good agreement
with the results reported in Tab.~\ref{tab_iter}.

Since the material evaluations for each
integration point are fully independent of another,
a parallelization
is easily implementable.
A dual thread material evaluation
reduced the simulation time to \SI{38.3}{min}
and thus almost halved the time difference.
For coarser meshes the
speed-up was lower due to the necessarily
introduced overhead of the parallelized implementation.

Under idealized circumstances the runtime can be reduced to $\mathcal{O}(N_tN_\text{Newton})$
and thus a constant in $N_\text{int}^\text{Fe}$ by performing all $N_\text{int}^\text{Fe}$
material evaluations on distinct threads.
In this example the additional simulation time would be only $\SI{55}{ms}$.

\section{Conclusion}
In this paper, we have presented a new stable
and quickly evaluable inversion scheme of the energy-based hysteresis model.
The static ferromagnetic model has been extended by the thin sheet eddy-current description
to obtain a dynamic hysteresis model.
It enabled an accurate simulation of a fast-ramping accelerator magnet
for sinusoidal and triangular pulsed current excitations.
The resulting differences in fields and loss as compared to
a static anhysteretic model were comparably small, granted correctly adjusted loss parameters
of the a posteriori formulae.
However, the dynamic hysteresis model is the only analyzed option
actively incorporating eddy current and hysteresis loss in the time domain simulation.
This property is crucial when analyzing the behavior of switched resonance circuits
consisting of capacitor banks and magnets.
We have shown that the added computational complexity in terms of material evaluation time is manageable and
can even on standard machines be reduced by exhausting parallelization.


%

%

\section*{Acknowledgment}
This work is funded by the European Union (EU) within the Horizon Europe Framework Programme (Project MuCol, grant agreement 101094300).
Views and opinions expressed are however those of the author(s) only and do not necessarily reflect those of the EU or European Research Executive Agency (REA).
Neither the EU nor the REA can be held responsible for them.
Endorsed by the IMCC.

\ifCLASSOPTIONcaptionsoff
  \newpage
\fi

\end{document}